\documentclass[12pt]{article}
\usepackage{amssymb}
\usepackage{epsfig}

\catcode`\@=11
\def\numberbysection{\@addtoreset{equation}{section}
        \def\theequation{\thesection.\arabic{equation}}}

\def\beq{\begin{equation}}
\def\eeq{\end{equation}}
\numberbysection
\begin{document}
\begin{titlepage}
\begin{center}
\hfill  \\
\vskip 1.in {\Large \bf Scalar field conformally coupled to a charged $BTZ$ black hole} \vskip 0.5in P. Valtancoli
\\[.2in]
{\em Dipartimento di Fisica, Polo Scientifico Universit\'a di Firenze \\
and INFN, Sezione di Firenze (Italy)\\
Via G. Sansone 1, 50019 Sesto Fiorentino, Italy}
\end{center}
\vskip .5in
\begin{abstract}
We study the Klein-Gordon equation of a scalar field conformally coupled to a charged $BTZ$ black hole. The background metric is obtained by coupling a non-linear and conformal invariant Maxwell field to $(2+1)$ gravity. We show that the radial part is generally solved by a Heun function and, in the pure gravity limit, by a hypergeometric function.
\end{abstract}
\medskip
\end{titlepage}
\pagenumbering{arabic}
\section{Introduction}

The knowledge of scalar fields in black hole backgrounds is useful for studying the physics of these objects. Once that the solutions of the Klein-Gordon equation for scalar fields are known, their properties allow to describe physical phenomena as the Hawking radiation of scalar particles.

The present article deals with a scalar field propagating in the background of $(2+1)$ gravity black holes, generalizing a previous paper dedicated to $AdS$ black holes of
$(3+1)$ gravity \cite{1}. For the sake of completeness we have explored a more general $(2+1)$ black hole solution endowed with electric charge.

Unfortunately the ( linear ) Maxwell action is not conformal invariant for dimension $d \neq 4$, and we have been forced to choose a non-linear generalization of the Maxwell action, dependent on a parameter $s$, introduced in \cite{2}-\cite{3}-\cite{4}. For a particular choice of the parameter $s$ ( $s = \frac{d}{4} $ ), the non-linear Maxwell action is indeed conformal invariant ( in our case $s = \frac{3}{4}$ ).

Once chosen the background metric we have studied the Klein-Gordon equation with conformal invariant coupling to the metric of the charged black hole. This special coupling allows us to simplify the solution. We are going to show that the radial solution can be expressed in terms of a Heun function with simple coefficients, analogously to what computed in the
$(3+1)$ case. In the pure gravity limit the black hole charge is null and the radial part reduces to a hypergeometric function, whose analytic properties are considerably simpler than the
Heun function.

This article is organized as follows. First we introduce the metric of $(2+1)$ black hole with non-linear Maxwell term and recall its thermodynamical properties.
Then we study in detail the solution of the Klein-Gordon equation in this background, simplifying the derivation of the radial part. We finally obtain a closed-form expression for the radial part in the pure gravity limit.

\section{The background metric}

In \cite{5} the metric of $( 2 + 1 ) $ gravity black hole has been introduced, with properties analogous to the physically relevant case in $( 3 + 1 ) $ dimensions.
However there are also some unphysical features due to the wrong number of dimensions. For example the $( 2 + 1 )$ gravity black hole admits closed temporal curves and its Hawking temperature is proportional to the mass parameter $M$, instead of being proportional to the inverse of it as it should be.

Despite these differences we believe that it still interesting discussing its properties since the $( 2 + 1 )$ gravity is a well known solvable system, and one expects to find simplifications that do not occur in $( 3 + 1 )$ dimensions. For a better comparison with the paper \cite{1}, we have studied a simple generalization of  the $( 2 + 1 )$ black hole
with the introduction of the electric charge. However a standard Maxwell term, being not conformal invariant, is not well suited for our method, and we have preferred discussing the Einstein gravity in presence of a non-linear Maxwell term \cite{2}-\cite{3}-\cite{4}:

\beq I ( g_{\mu\nu} , A_{\mu} ) \ = \ \frac{ 1 }{ 16 \pi G} \ \int \ d^3 x \ \sqrt{|g|} \ [  \ R - 2 \Lambda + ( k F )^s \ ] \ \ \ \ \ \ s >  \frac{ 1 }{ 2 } \label{21}\eeq

where the symbol $ F = F^{\mu\nu} \ F_{\mu\nu} $  stands for the Maxwell invariant. The restriction to $s > \frac{ 1 }{ 2 }$ is required in order to obtain an asymptotically well-defined electric field.

By varying this action with respect to the metric tensor $ g_{\mu\nu} $ and the electromagnetic field $ A_{\mu} $, the equations of motion for the gravitational and electromagnetic fields
can be derived as follows

\begin{eqnarray}
G_{\mu\nu} + \Lambda  g_{\mu\nu} & = & T_{\mu\nu} \nonumber \\
\partial_{\mu} \ ( \ \sqrt{|g|} F^{\mu\nu} ( k F )^{s-1} \ ) & = & 0 \label{22}
\end{eqnarray}

where the energy-momentum tensor is given by

\beq T_{\mu\nu}  \ = \ 2 [ \  s k F_{\mu\rho} F^{\rho}_{\nu} ( k F )^{s-1} \ - \ \frac{ 1 }{ 4 } g_{\mu\nu} ( k F )^s \ ] \label{23} \eeq

The choice  $s=1$ and $k =-1$ faithfully reproduces the ordinary Einstein-Maxwell equations of motion.

In the non-linear case by choosing $k =-1$  we obtain a novel solution for the charged black hole. The electromagnetic field is  given by

\beq
F_{tr} \ = \ \left\{ \begin{array}{cc}  \frac{q}{r} & \mbox{s=1}  \\
         - \frac{ 2q (s-1) }{ 2s-1 } \ r^{-\frac{ 1 }{ 2s -1 }} & \mbox{otherwise} \end{array} \right. \label{24} \eeq

The corresponding metric is

\beq ds^2 \ = \ g(r) dt^2 \ - \ \frac{ dr^2 }{ g(r) } \ - \ r^2 d \phi^2 \label{25} \eeq

with the function $g(r)$ given by

\beq g(r) \ = \ |\Lambda| r^2 \ - \ M \  + \ \left\{ \begin{array}{cc}  - \ 2 q^2 ln ( \sqrt{ |\Lambda| } r ) & \mbox{s=1}  \\
\frac{ (2s-1)^2 }{ 2(s-1) } \ \left( \frac{ 8 q^2 (s-1)^2 }{ (2s-1)^2 } \right)^s \ r^{\frac{ 2(s-1) }{ 2s-1 }}  & \mbox{otherwise} \end{array} \right. \label{26} \eeq

where $q$ is the electric charge of the black hole.

Now we can specialize to the case $s = \frac{ 3 }{ 4 }$, the so-called conformal invariant Maxwell field ( in general $s = \frac{ d }{ 4 }$ with $d$ the space-time dimensions ).
In this case the electric field behaves as $r^{-2}$ as it happens to the Maxwell field in $(3+1)$ dimensions:

\begin{eqnarray}
F_{tr} & = &   \frac{ q }{ r^2 }\nonumber \\
g(r)   & = &  |\Lambda| r^2 \ - \ M \ - \ \frac{ (2q^2)^\frac{ 3 }{ 4 } }{ 2r } \label{27}
\end{eqnarray}

The metric function $g(r)$ has a positive real root, as it happens in the case without electric charge. By choosing the constant $K$ as

\beq K \ = \ ( 2 q^2 )^\frac{ 3 }{ 4 } \label{28} \eeq

we obtain the definitive version of the black hole metric which we will study in the following sections:

\beq ds^2 \ = \ \left( |\Lambda| r^2 \ - \ M \ - \ \frac{ K }{ 2r } \right) dt^2 \ - \ \frac{ dr^2 }{ ( |\Lambda| r^2 \ - \ M \ - \ \frac{ K }{ 2r } ) } \ - \ r^2 d \phi^2
\label{29} \eeq

The metric is stationary and with axial symmetry.

There is a special value in which the metric function $g(r)$ admits three real roots

\beq K^2 \ = \ \frac{ 16 }{ 27 } \ \frac{ M^3 }{ |\Lambda| } \label{210} \eeq

In this case the horizon $r_1$  and the other two roots $r_2, r_3$ are situated in

\beq r_1^2 \ = \ \frac{ 4 }{ 3 } \ \frac{ M }{ |\Lambda| } \ \ \ \ \ \ \ r_2 \ = \ r_3 \ = - \ \frac{ r_1 }{ 2 } \label{211} \eeq

We note that outside the singularity $r=0$ the following property holds

\beq R \ = \ - \ 6 \ |\Lambda| \ \ \ \ \ \ \ \ s = \frac{ 3 }{ 4 } \label{212} \eeq

since the energy-momentum tensor is traceless

\beq T^\mu_\mu = 0 \label{213} \eeq

As far as the thermodynamical properties are concerned, we know that the Hawking temperature is generally given by

\beq
T_1 \ = \ \left\{ \begin{array}{cc}  \frac{|\Lambda|}{ 2\pi } \left( r_1 \ - \ \frac{ q^2 }{ |\Lambda| r_1 }
\right)  & \mbox{s=1}  \\
         \frac{|\Lambda|}{ 4\pi } \ \left( 2 r_1 \ + \ \frac{ 2s-1 }{ |\Lambda| } \ \left( \frac{ 8 q^2 (s-1)^2 }{ (2s-1)^2 } \right)^s \ r^{\frac{1 }{ 1-2s }}_1
\right)  & \mbox{otherwise} \end{array} \right. \label{214} \eeq

In the particular case $ s = \frac{ 3 }{ 4 } $

\beq  T_1 \ = \ \frac{|\Lambda|}{ 4\pi } \left( 2 r_1 \ + \ \frac{ K }{ 2 |\Lambda| r_1^2 }
\right)  \label{215} \eeq

\section{Equation of motion for the scalar field}

We are going to discuss a ( neutral ) scalar field propagating in the metric of the charged $BTZ$ black hole given by (\ref{29}). Obviously we omit the scalar field energy-momentum tensor in the Einstein equations for the black hole. We know from the analysis performed in $(3+1)$ dimensions, that the equation of motion for the scalar field simplifies when we choose a conformal invariant coupling with the gravitational field. The action of a $(2+1)$ dimensional scalar field conformally coupled to the background metric (\ref{29}), introduced in the previous section, is given by

\beq S \ = \ \int \ d^D x \ \sqrt{|g|} \ \frac{ 1 }{ 2 } \ \left( \ g^{\mu\nu} \ \partial_{\mu} \phi \
\partial_{\nu} \phi \ - \ \frac{1}{8} \ R \ \phi^2 \ \right) \label{31}\eeq

The corresponding equation of motion is

\beq ( \ \Box \ + \ \frac{1}{8} \ R \ ) \ \phi \ = \ 0 \ \ \ \ \ \ \ \ \ \Box \ = \ |g|^{ - \frac{ 1 }{ 2 } } \
\partial_\mu \ ( \ |g|^{ \frac{ 1 }{ 2 } } \ g^{ \mu \nu } \ \partial_\nu \ ) \label{32}\eeq

From the black hole metric (\ref{29}) we get the following components

\begin{eqnarray}
 & \ & g_{tt} \ = \ - \ g^{rr} \ = \ \frac{Q(r)}{r^2} \ \ \ \ \ \ \ g^{tt} \ = \ - \ g_{rr} \ = \ \frac{r^2}{Q(r)}
 \ \nonumber \\
 & \ & Q(r) \ = \ r \left[ |\Lambda| r^3 \ - \ M r \ - \ \frac{K}{2} \right] \ = \ |\Lambda| r ( r - r_1 ) ( r - r_2 ) ( r - r_3 ) \nonumber \\
 & \ & R \ = \ - \  6 \ |\Lambda| \label{33}
\end{eqnarray}

and the equation of motion for the scalar field reads now

\beq \frac{r}{Q(r)} \partial_r \left( \frac{Q(r)}{r} \partial_r \right) \phi \ - \ \frac{ r^4 }{ Q^2 (r) } \ \partial_t^2 \phi \ + \ \frac{ 1 }{ Q (r) } \ \partial_\theta^2 \phi \ + \ \frac{3}{4} \ | \Lambda |  \ \frac{r^2}{Q(r)} \phi \ = \ 0 \label{34}
\eeq

This equation can be solved by splitting the variables

\begin{eqnarray} & \ & \phi \ = \ R(r) \ e^{ i \lambda \phi } \ e^{ - i \omega t} \nonumber \\
 & \ & \partial_r^2 R(r) \ + \  \partial_r \ ln \left( \frac{Q(r)}{r} \right) \ \partial_r R(r) \ + \ \frac{ \omega^2 r^4 }{ Q^2 (r) } \ R(r) \ - \ \frac{ \lambda^2 }{ Q(r) } \ R(r) \ + \ \frac{ 3 }{ 4 } \ |\Lambda| \ \frac{ r^2 }{ Q(r) }  \ R(r) \ = \ 0 \nonumber \\
 & \ &  \label{35} \end{eqnarray}

We can map the roots $ r = 0 $ and $ r = r_1 $ of the equation $ Q(r) = 0 $ in the points $ 0, 1 $ through the following change of variable

\beq z \ = \ \frac{ r ( r_1 - r_3 ) }{ ( r - r_3 ) r_1 } \label{36} \eeq

This transformation maps the other two roots $r_2, r_3$  in the points

\beq r_2 \rightarrow z = \xi = \frac{ r_2 ( r_1 - r_3 ) }{ ( r_2 - r_3 ) r_1 } \ \ \ \ \ r_3 \rightarrow z = \infty \label{37} \eeq

while the singularity at infinity $ r = \infty $ is mapped into the point

\beq r \ = \ \infty \ \rightarrow \ z = \eta = \frac{ r_1 - r_3 }{ r_1 } \label{38} \eeq

By using the following identity

\beq \left( \ \frac{ 1 }{ r }, \ \frac{ 1 }{ r - r_1 }, \ \frac{ 1 }{ r - r_2 }, \ \frac{ 1 }{ r - r_3 } \
\right) \ = \ \frac{ z - \eta }{ \eta \ r_3 } \ \left( \ \frac{ \eta }{ z }, \ \frac{ \eta - 1 }{ z - 1 },
\ \frac{ \eta - \xi }{ z - \xi }, \ 1 \right) \label{39} \eeq

the radial equation can be rewritten in the $z$ variable as follows

\begin{eqnarray}
  \frac{ d^2 }{ d z^2 } \ R(z) & + & \ P(z) \ \frac{ d }{ d z } R(z) \ + \ Q(z) \  R(z) \ = \ 0 \nonumber \\
  P(z) & = & \frac{ 1 }{ z - 1 } \ + \ \frac{ 1 }{ z - \xi } \ - \ \frac{ 1 }{ z - \eta }
 \nonumber \\
 Q(z) & = & \frac{ \omega^2_1 }{ ( z - 1 )^2 } \ + \ \frac{ \omega^2_2 }{ ( z - \xi )^2 } \ + \ \frac{ 2 \omega_1 \omega_2 }{ ( z - 1 ) \ ( z - \xi ) } \nonumber \\
 & + & \frac{ 3 }{ 4 } \ \frac{ 1 }{  \ ( z - \eta )^2 } \ - \ \frac{ 3 }{ 2 \ \eta } \ \frac{ 1 }{ ( z - \eta ) } \ - \ \frac{ 3  }{ 2 \ \eta }
 \frac{ 1 + \xi - z }{ ( z -1 ) ( z - \xi ) } \ + \ \frac{ 3 }{ 4 } \ \frac{ 1 }{  \ ( z -1 ) ( z - \xi ) } \nonumber \\
 & - & \frac{ \lambda^2 }{ |\Lambda| \ r_1 ( r_2 - r_3 ) } \ \frac{ 1 }{ z ( z - 1 ) ( z - \xi ) } \label{310}
\end{eqnarray}

where we define the parameters

\beq  \omega_1 = \frac{\omega }{ |\Lambda| \  \eta \ ( r_1 - r_2 )  } \ \ \ \ \ \ \ \ \omega_2 = - \ \xi \ \omega_1
\label{311} \eeq

Analogously to the $(3+1)$ case we find the complete removal of the singularity in $(z-\eta)$ once that the radial equation is put into its normal form

\begin{eqnarray}
R(z) & = & e^{-\frac{ 1 }{ 2 } \int P(z) \  dz} \ Z(z) \nonumber \\
\frac{ d^2 }{ dz^2 } Z(z) & + & \tilde{ Q }(z) \ Z(z) = 0 \label{312}
\end{eqnarray}

where we define

\begin{eqnarray}
\tilde{ Q }(z) & = & Q(z) \ - \ \frac{ 1 }{ 2 } \ \frac{ d }{ dz } P(z) \ - \ \frac{ 1 }{ 4 } \ P^2(z) = \nonumber \\
& = &  \left( \frac{ 1 }{ 4 } \ + \ \omega_1^2 \right) \ \frac{ 1 }{ ( z-1 )^2 } \ + \
 \left( \frac{ 1 }{ 4 } \ + \ \omega_2^2 \right) \ \frac{ 1 }{ ( z-\xi )^2 } \ - \ \left( \frac{ 1 }{ 4 } \ - \ 2 \omega_1 \omega_2 \right) \ \frac{ 1 }{
(z-1) ( z-\xi ) } \nonumber \\
& - & \frac{ \lambda^2 }{ |\Lambda| \ r_1 ( r_2 - r_3 ) } \ \frac{ 1 }{ z ( z - 1 ) ( z - \xi ) } \label{313}
\end{eqnarray}

This equation is simply related to the Heun equation

\beq \frac{ d^2 }{ dz^2 } w(z) \ + \ P_H(z) \ \frac{ d }{ dz } w(z) \ + \ Q_H(z) w(z)  \ = \ 0 \label{314} \eeq

where

\begin{eqnarray}
P_H(z) & = & \frac{ \gamma }{ z } \ + \ \frac{ \delta }{ z-1 } \ + \ \frac{ \epsilon }{ z-\xi } \nonumber \\
Q_H(z) & = & \frac{ \alpha \beta z - q  }{ z ( z-1 ) ( z-\xi )} \label{315}
\end{eqnarray}

Once we rewrite the Heun equation into its normal form, we must identify it with equation (\ref{313})

\begin{eqnarray}
w(z) & = & e^{-\frac{ 1 }{ 2 } \int P_H(z) \ dz} \ Z(z) \nonumber \\
\tilde{ Q }(z) & = &  \frac{ \frac{\gamma}{2} ( 1 - \frac{\gamma}{2} ) }{ z^2 } \ + \
\frac{ \frac{\delta}{2} ( 1 - \frac{\delta}{2} )}{ ( z-1 )^2 } \ + \
\frac{ \frac{\epsilon}{2} ( 1 - \frac{\epsilon}{2} )}{ ( z-\xi )^2 } \nonumber \\
& + & \frac{ [ \alpha \beta - \frac{ 1 }{ 2 } ( \gamma \delta + \gamma \epsilon + \delta \epsilon ) ]
}{ (z-1) ( z-\xi ) } \ - \  \frac{ q - \frac{ 1 }{ 2 } ( \gamma \delta \xi + \gamma \epsilon ) }{ z ( z-1 ) ( z-\xi) } \label{316} \end{eqnarray}

We must also add the Heun condition

\beq \alpha \ + \ \beta \ = \ \gamma \ + \ \delta \ + \ \epsilon \ - \ 1 \label{317} \eeq

We find that fixing the values of

\beq \delta \ = \ 1 \ + \ 2i \ \omega_1 \ \ \ \ \ \epsilon \ = \ 1 \ + \ 2i \ \omega_2   \label{318} \eeq

there are two possible choices for the other parameters

\beq \alpha \ = \ \frac{ 1 }{ 2 } \ \ \ \ \ \ \beta \ = \ \frac{ 1 }{ 2 } \ + \ 2 i \ ( \ \omega_1 \ + \  \omega_2 \ )  \ \ \ \ \ \ \gamma \ = \ 0
\ \ \ \ \ \ q \ = \ \frac{ \lambda^2 }{ |\Lambda| \ r_1 ( r_2 - r_3 ) } \label{319} \eeq

or

\beq \alpha \ = \ \frac{ 3 }{ 2 } \ \ \ \ \ \ \beta \ = \ \frac{ 3 }{ 2 } \ + \ 2 i \ ( \ \omega_1 \ + \  \omega_2 \ )  \ \ \ \ \ \ \gamma \ = \ 2
\ \ \ \ \ \ q \ = \ \frac{ \lambda^2 }{ |\Lambda| \ r_1 ( r_2 - r_3 ) } \ + \ 1 + \xi \label{320} \eeq

In the special case (\ref{210}) we get the confluence of the singularity $ z = \xi $ with the singularity at infinity, but we omit it since it's not particularly interesting.
Instead in the following section we will study in detail the pure gravity case that gives rise to the confluence between the singularity $ z = \xi $ and $ z=0 $.

\section{Pure gravity limit}

In the pure gravity limit the black hole electric charge is null, the parameter $ K = 0 $ from which we derive the following consequences

\beq r_2 \rightarrow 0  \ \ \ \ \ \ r_1 \rightarrow - r_3 \ \ \ \ \ \ \ \ \xi \rightarrow 0 \ \ \ \ \ \ \ \ \eta \rightarrow 2 \label{41} \eeq

Therefore the radial equation in the $z$ variable reduces to

\begin{eqnarray}
  \frac{ d^2 }{ d z^2 } \ R(z) & + & \ P(z) \ \frac{ d }{ d z } R(z) \ + \ Q(z) \  R(z) \ = \ 0 \nonumber \\
  P(z) & = & \frac{ 1 }{ z } \ + \ \frac{ 1 }{ z - 1 } \ - \ \frac{ 1 }{ z - 2 }
 \nonumber \\
 Q(z) & = & \frac{ \omega^2_1 }{ ( z - 1 )^2 } \ + \ \frac{ 3 }{ 4 } \ \frac{ 1 }{  \ ( z - 2 )^2 } \ - \ \frac{ 3 }{ 4 } \ \frac{ 1 }{ ( z - 2 ) } \ + \ \frac{ 3 }{ 4 } \
 \frac{ 1 }{ z -1 } \nonumber \\
 & - & \frac{ \lambda^2 }{ M } \ \frac{ 1 }{ z^2 ( z - 1 )} \label{42}
\end{eqnarray}

and going into the normal form

\begin{eqnarray}
R(z) & = & e^{-\frac{ 1 }{ 2 } \int P(z) \  dz} \ Z(z) \nonumber \\
\frac{ d^2 }{ dz^2 } Z(z) & + & \tilde{ Q }(z) \ Z(z) = 0 \label{43}
\end{eqnarray}

we get

\beq
\tilde{ Q }(z) \ = \ \left( \frac{ 1 }{ 4 } \ + \ \frac{ \lambda^2 }{ M } \right) \ \frac{ 1 }{ z^2 } \ + \
\left( \frac{ 1 }{ 4 } \ + \ \omega_1^2 \right) \ \frac{ 1 }{ ( z-1 )^2 } \ - \ \left( \frac{ 1 }{ 4 } \ + \ \frac{ \lambda^2 }{ M } \right) \ \frac{ 1 }{
z (z-1) } \label{44}
\eeq

This normal form is equivalent to an hypergeometric function with coincident critical exponents at the infinity and at the horizon ( $z=1$ ).

The solution for the radial part can be expressed as follows

\beq R(z) \ = \ \sqrt{\frac{ z - 2 }{ z ( z-1 ) }} \ Z(z) \ \propto \ z^{ \beta - \frac{ 1 }{ 2 }  } \ ( z-1 )^{ i \omega_1 } \ ( z- 2 )^{ \frac{ 1 }{ 2 }  } \ {}_2 F_1 ( \beta + 2 i \omega_1 , \beta, 2 \beta ; z ) \label{45}
\eeq

where we have introduced the parameter $\beta$

\beq \beta \ = \ \frac{ 1 }{ 2 } \ + \ i \frac{ \lambda }{ \sqrt{M} } \label{46}
\eeq

By using the identity

\beq {}_2 F_1 ( \alpha , \beta, 2 \beta ; z ) \ = \ \left( 1- \frac{ z }{ 2 } \right)^{-\alpha} \ {}_2 F_1 \left( \frac{  \alpha }{ 2 } , \frac{ \alpha + 1 }{ 2 } , \beta + \frac{ 1 }{ 2 } ; \left( \frac{ z }{ 2-z } \right)^2 \right) \label{47}
\eeq

and going back to the original radial variable $r$ we arrive at the final formula

\begin{eqnarray}
R(r) & \propto & \left( \frac{ r^2 - r^2_1 }{ r^2_1 }\right)^{ i \omega_1 } \ \left( \frac{ r }{ r_1 }\right)^{ i \frac{ \lambda }{ \sqrt{M} } } \
{}_2 F_1 \left( \frac{ 1 }{ 4 } + \frac{ i }{ 2 } \frac{ \lambda }{ \sqrt{M} } + i \omega_1 , \frac{ 3 }{ 4 } + \frac{ i }{ 2 } \frac{ \lambda }{ \sqrt{M} } + i \omega_1  ,
1 + i \frac{ \lambda }{ \sqrt{M} } ; \left( \frac{ r }{ r_1 } \right)^2 \right) \nonumber \\
& \ & \label{48}
\end{eqnarray}

Taking into account the symmetry of the differential equation (\ref{44}), all the independent solutions are generated by exchanging the parameters $ \lambda \rightarrow - \lambda$ and $ \omega_1 \rightarrow - \omega_1 $.

\section{Conclusions}

In this article we have presented analytical solutions for the Klein-Gordon equation of a scalar field conformally coupled to a charged $BTZ$ black hole.
The background metric is obtained by coupling a non-linear and conformal invariant Maxwell field to $(2 +1)$ gravity. These solutions are analytical in the whole region between the event horizon and the infinity. The solution for the radial part is given in terms of a Heun function and, in the pure gravity limit, in terms of a hypergeometric function.
The knowledge of such solutions ( thanks to their analytical properties ) allows an exhaustive analysis of the Hawking radiation effect without the need of introducing the tortoise coordinate, which is valid only locally.

\end{document}